\documentclass[preprint]{revtex4}
\usepackage{graphicx}
\usepackage{subfigure}
\usepackage{epsfig}
\usepackage{amssymb}
\usepackage{amsmath}

\begin{document}
                                                                                
\newcommand{\lP}{l_{\mathrm P}}
                                                                                
\newcommand{\md}{{\mathrm{d}}}
\newcommand{\tr}{\mbox{tr}}
                                                                                
\newcommand*{\R}{{\mathbb R}}
\newcommand*{\N}{{\mathbb N}}
\newcommand*{\Z}{{\mathbb Z}}
\newcommand*{\Q}{{\mathbb Q}}
\newcommand*{\C}{{\mathbb C}}

\newcommand{\ket}[1]{| #1 \rangle}
\newcommand{\bra}[1]{\left\langle #1 \right|}
\newcommand{\braket}[2]{\langle #1 \,|\, #2 \rangle }
\newcommand{\braketfull}[3]{\langle #1 \,|\, #2 \,|\, #3 \rangle }

\newcommand{\bas}{\ket{\mu}}
\newcommand{\half}{\frac{1}{2}}
\newcommand{\sgn}{\text{sgn}}
\newcommand{\cof}[1]{s_1(\mu #1) \, \psi_{\mu #1}}
\newcommand{\cs}{\widehat{\text{cs}}}
\newcommand{\sn}{\widehat{\text{sn}}}

\title{On the Hamiltonian Constraint of Loop Quantum Cosmology}
\author{Kevin Vandersloot}
\email{kfvander@gravity.psu.edu}
\affiliation{Institute for Gravitational Physics and Geometry, The Pennsylvania
State University, 104 Davey Lab, University Park, PA 16802, USA}

\begin{abstract}
In this paper we construct the Hamiltonian constraint operator
of loop quantum cosmology using holonomies defined for arbitrary
irreducible SU(2) representations labeled by spin $J$. We
show that modifications to the effective semi-classical equations of motion arise both
in the gravitational part of the constraint as well as matter terms.
The modifications are important for phenomenological
investigations of the cosmological imprints of
loop quantum cosmology. We discuss the implications
for the early universe evolution.
\end{abstract}

\maketitle

\section{Introduction}
In the study of cosmology many questions remain unanswered
despite the wealth of observational data at hand. In particular
our understanding of the early universe is lacking. It is
at these high energies that any classical cosmological
models are no longer relevant and a quantum
theory of gravity is needed. A deep quantum understanding
of the regime is required to answer, for instance,
questions related to the initial big-bang singularity
and to understand what conditions preceded the onset of chaotic
inflation.

A particularly exciting quantum theory of cosmology is known as loop quantum
cosmology (LQC) (for a review and references see \cite{Bojowald:2004ax}). The progress made in LQC to date
has been impressive. LQC has provided explanations to long standing questions in quantum
cosmology. The reason for its success can be traced to the fact that it has as its basis a
candidate full theory of quantum gravity known as loop quantum gravity (for reviews see
\cite{book, ash10, Thiemann:2002nj, lee1,ap}). Thus LQC is expected to handle on a firmer basis the extreme high energy regime
of the early universe where non-perturbative effects are paramount. For instance it has
been shown that the theory contains a robust mechanism for singularity 
avoidance \cite{Bojowald:2001xe}. The
mechanism arises as a natural feature of the theory and does not rely on ad-hoc forms of
matter. This is in contrast to earlier forms of quantum cosmology such as those based on
Wheeler-DeWitt quantization where singularity avoidance does not arise naturally.

Yet, an even more exciting result is that LQC predicts modifications to the evolution of
the universe for scales larger than the Planck scale. For instance it has been shown that
LQC can modify the behavior of the inverse volume such that it becomes a bounded function
\cite{Bojowald:2001vw}. This gives rise to modifications to the behavior of matter in the form of a scalar field
which in turn predicts a period of super-inflation at energies near the Planck scale \cite{Bojowald:2002nz}. In
this super-inflationary regime the inflaton is pushed up its potential due to the existence of
an anti-friction term in the Klein-Gordon equation which can set the stage for the onset of
slow-roll inflation \cite{Bojowald:2003mc, Tsujikawa:2003vr}. The modified dynamics in this regime can even lead to measurable signatures at large scales in the CMB \cite{Tsujikawa:2003vr}.

Most of the phenomenological studies to date have exploited the modification of the
inverse volume. For instance, in a collapsing universe with a scalar field the anti-frictional
term in the Klein-Gordon equation becomes frictional and a bounce can occur \cite{Singh:2003au}. This
effect has been studied for a scalar field with a self-interaction potential in positive curvature
oscillatory universes where the effects of repeated expansion/contraction push the scalar field
up its potential providing another mechanism to establish the initial conditions for slow-roll inflation \cite{Mulryne:2004va, Lidsey:2004ef}. In
addition the modifications can lead to a bounce in the context of cyclic models with colliding
branes, possibly removing singular behavior that has plagued these models \cite{Bojowald:2004kt}.

These results are intriguing yet some open questions remain as to their
validity.  The modifications to the inverse volume arise 
by exploiting an ambiguity in defining the inverse volume operator
which is required to quantize the matter part of the Hamiltonian constraint
for a scalar field. In defining the inverse volume operator one traces
over SU(2) valued holonomies and the ambiguity lies in choosing
the irreducible representation in which to perform the trace. The 
ambiguity is labeled by a half-integer $J$ representing
the spin of the representation. A neglected fact
is that this same ambiguity arises in the gravitational part of
the Hamiltonian constraint where one also traces over
SU(2) valued holonomies. So far, the gravitational part of the constraint
has been quantized using the $J=1/2$ representation while
the ambiguity has been freely exploited for the matter part. 

It is therefore not clear if corrections arising in the gravitational
part might change the dynamics and alter significantly the phenomenological
consequences. From the full theory the gravitational part contains an inverse volume factor
and the possibility remains that modifications here might cancel the effects in the matter
part. It is thus of critical importance to determine the complete quantum corrections for
arbitrary $J$ arising in both terms of the Hamiltonian constraint.

In this paper we aim to clear up the uncertainty with respect to this ambiguity. We
explicitly construct the isotropic Hamiltonian constraint operator for arbitrary spin $J$ representations. We show that the resulting difference equation is higher order for larger $J$
and that geometric quantities in the gravitational part of the constraint obtain quantum modifications analogously to the inverse volume modifications for
the matter part. 
We show that the higher $J$ constraint operator contains the
same singularity resolution mechanism attesting to its robustness. 
We will derive and motivate effective semi-classical equations of motion
from the constraint operator and comment  on the phenomenological consequences 
and show that the qualitative results do not change. 
As an explicit
example, the $J=1$ constraint operator is constructed and analyzed with special attention
payed to the semi-classical limit.

\section{Loop Quantum Cosmology}
We start with a brief overview of LQC and its kinematical Hilbert space. In Dirac
quantization for systems with constraints (such as general relativity) one first starts by
constructing a kinematical Hilbert space by quantizing the system ignoring the constraints.
The constraints are then represented as self-adjoint operators on this space and physical
solutions are wave functions annihilated by the constraint operators. In this paper we will
restrict ourselves to homogeneous and isotropic models with zero spatial curvature. Upon
the imposition of this symmetry reduction, the Hamiltonian of isotropic GR consists of a
single constraint, the Hamiltonian constraint. The details
of the construction of the kinematical Hilbert space and
the constraint operator are given in \cite{Ashtekar:2003hd}. We will
follow the same conventions and notation here.

The first step toward quantization is a classical symmetry reduction of the action of
general relativity written in connection form. In the full theory of loop quantum gravity
the action is written in terms of a connection $A$ which contains the information about
curvature, as well as a triad $E$ conjugate to $A$ which encodes the geometry. After reduction
to homogeneity and isotropy the connection and triad are labeled by single quantities $c$ and
$p$ respectively. These variables satisfy the commutation relation
\begin{equation*}
\{c,p\} = \frac{1}{3} \kappa \gamma
\end{equation*}
where $\kappa = 8\pi G$ and $\gamma$ is the Barber-Immirzi parameter
which represents a quantum ambiguity (black hole entropy
calculations can fix its value to $\gamma \approx .2735$ 
\cite{Meissner:2004ju,Domagala:2004jt} ) .
In terms of the standard metric variables given by the scale factor
$a$ , the new variables are related as
\begin{subequations}
\begin{eqnarray}
	|p| &=& a^2 \label{p}\\
	c &=& \gamma \dot{a} \label{c}
\end{eqnarray}
\end{subequations}
Note the absence of the factor of $1/2$ in the formula for
the connection (\ref{c}) which appears in \cite{Bojowald:2002nz}. The correct
factor (without the $1/2$) can be derived using the relation
from the full theory $A^i_a = \Gamma^i_a + \gamma K^i_a$
where $A^i_a$ is the full connection, $K^i_a$ is the extrinsic
curvature, and $\Gamma^i_a$ is the spin connections which
vanishes identically for isotropic flat models. The absolute value
around the triad component $p$ indicates that we are extending
the classical phase space of isotropic GR to include both orientations
of the triad (positive and negative $p$). Note that this does not
mean that negative $p$ corresponds to a negative scalar factor. For both
regions the scale factor, and hence the volume, is positive.

Dynamics is determined completely by the Hamiltonian constraint
$H$ given by
\begin{equation}\label{Hclass}
	H = - \frac{3}{\kappa \gamma^2} \sgn(p) \sqrt{|p|} \; c^2
	+ H_{m}
\end{equation}
where $H_m$ is the matter Hamiltonian.
Note again the factor of $3$ which differs from previous works. With
the correct factors here the resulting action
$S_{GR} = \int \! dt \frac{3}{\kappa \gamma} p \dot{c} \,-\, N H_{GR}$
is equivalent to the standard isotropic Einstein-Hilbert action
$S_{GR} = \frac{1}{2\kappa} \int \! \sqrt{|g|}\, R$, and the
Hamiltonian equations $\dot{p} = \{p,H\}$ and 
$\dot{c} = \{c,H\}$ are equivalent to the equations of
motion derived from the Einstein-Hilbert action.

With the symmetry reduced action we now turn to quantization.
The theory is quantized using ideas and techniques from the full theory
of loop quantum gravity. The basic variables to be quantized are holonomies of
the connection and the smeared triad integrated on a two-surface.
Using this, the quantum configuration space inherits a discrete structure
and an {\em orthonormal} basis is given by states labeled by a parameter
$\mu$ given by
\begin{equation*}
	\braket{c}{\mu} = \exp(i \mu c /2) \;\;\;\;\;\; \mu \in \R \,.
\end{equation*}
The basis states $\bas$ are eigenstates of the triad operator
$\hat{p}$ with eigenvalues
\begin{equation}\label{pmu}
	\hat{p} \;\bas = \frac{\mu \gamma l_p^2}{6} \;\bas \,.
\end{equation}
Geometric operators are built from the triad, which
allows us to understand the physical meaning of the parameter $\mu$.
In particular it is an eigenstate of the volume operator
$\widehat{V} = |\hat{p}|^{3/2}$ hence the physical volume
is given by
\begin{equation}
	V_{\mu} = \left( \left| \frac{\gamma \mu l_p^2}{6} \right| \right)^{3/2}\,.
\end{equation}

More complicated operators are built from the basic triad operator
$\hat{p}$ and holonomy operators. Because of homogeneity and isotropy we
only need consider holonomies along straight lines which can be
parameterized as
\begin{equation*}
	h_i = \exp(\mu_0 c \tau_i)
\end{equation*}
where $\tau_i$ are the generators of the Lie algebra of SU(2)
satisfying $[\tau_i,\tau_j] = \epsilon^k_{ij} \tau_k$, and $\mu_0$ is
proportional to the length of the holonomy. In the fundamental representation the
holonomies are given by the simple formula
\begin{equation*}
	^{(\frac{1}{2})}h_i = \cos(\mu_0 c / 2) + 2 \tau_i \sin( \mu_0 c / 2)
\end{equation*}
where the notation $^{(\frac{1}{2})}$ indicates that the holonomy
is defined in the fundamental representation ($J=1/2$). In this
paper we will use superscripts to the left of an object to
indicate the irreducible representation of SU(2) used.
The holonomy operators in turn
behave as exponentiated position operators and thus act
on the basis states by finite shifts. More precisely, the holonomy
operators consist of the cosine and sine
operators which we will for compactness notate
$\cs$ and $sn$ with actions
\begin{eqnarray}\label{hops}
       \cs \;\bas &\equiv& \cos\left( \frac{\mu_0 \hat{c}}{2} \right) \bas
        = \frac{1}{2} \Big[ \ket{\mu+\mu_0}+ \ket{\mu-\mu_0} \Big]
	\nonumber \\
	\sn\;\bas &\equiv& \sin\left( \frac{\mu_0 \hat{c}}{2} \right) \bas
	 = -\frac{i}{2} \Big[ \ket{\mu+\mu_0}- \ket{\mu-\mu_0} \Big]
\end{eqnarray}

Knowing the action of the holonomy operators let us now turn to 
two essential operators: the inverse volume and the gravitational
part of the Hamiltonian constraint. The inverse volume operator
is needed to quantize certain forms of matter, for instance
a scalar field. Since the eigenstates of the volume operator are 
normalizable, the naive inverse volume operator (with
eigenvalues equal to the inverse of the volume eigenvalues) is
not a densely designed self-adjoint operator. The solution is to recast the classical
formula for the inverse volume using the Poisson bracket between the connection and the volume
\cite{Thiemann:1996aw}. When quantized the connection is represented with 
a holonomy operator, and the Poisson bracket is replaced with a commutator.
The resulting operator is diagonal in the $\bas$ basis and is given
by \cite{Ashtekar:2003hd}
\begin{equation}\label{dj1/2}
	^{(\half)} \widehat{V^{-1}}\; \bas =  \left[ \frac{4}{\gamma l_p^2}
	\left( V_{\mu+1}^{1/2} - V_{\mu-1}^{1/2} \right) \right]^6 \bas \,.
\end{equation}
The eigenvalues are in fact bounded and approach zero near the classical
singularity $\mu=0$. The fact that the inverse volume is cutoff for small
volumes is the principal reason for the period of super-inflation for
a scalar field in the early universe. We will discuss this more later.

There are two important points we wish to make regarding the
inverse volume operator given here. First, as noted the formula
is obtained using traces of holonomies in the
spin $1/2$ representation. The generalization to higher  $J$
has been performed \cite{Bojowald:2002ny} and the
effect is that the cutoff region gets pushed to larger volumes for
larger $J$ allowing for phenomenological modifications for scales larger
than the Planck length (the cutoff for $J=1/2$ is deep in the Planckian regime where
semi-classical equations of motion are no longer expected to be valid). Most
of the phenomenological investigations are based on
the modifications for larger values of $J$. The second point we make
is that the holonomy operators (\ref{hops}) involve the
length parameter $\mu_0$ and an implicit value
of $\mu_0=1$ has been used to arrive at equation (\ref{dj1/2}). As
stated in \cite{Ashtekar:2003hd} one can use an arbitrary value $\mu_0$ in
the regularization and in that reference the Hamiltonian constraint is
constructed using a value of $\sqrt{3}/4$ determined on physical
grounds from the smallest allowed area of the full theory of LQG. 
It is natural that the same
value should be used both for the Hamiltonian constraint and the
inverse volume operator since both are regularized using holonomies.
We would like to update the formulas for the inverse volume operator to
account for this ambiguity and in the next section we will do so.

Finally we wish to represent the Hamiltonian constraint operator,
classically given in equation (\ref{Hclass}). There are two important non-trivialities that arise when attempting to quantize the constraint. The first is that the connection parameter $c$ is no longer one of the basic
variables and must be represented using the holonomy operators. The other important non-triviality deals with how the $\sgn(p) \sqrt{p}$ term of the constraint gets quantized from the full
theory. While this term in the isotropic constraint could be simply quantized using the 
$p$ operator (\ref{pmu}), this approach is not possible in the full theory. There the full constraint is
given by \cite{ash10}
\begin{equation*}
	H = \frac{1}{\kappa} \int \!\! d^3x \;N\; \epsilon^{ij}_k \;
	\frac{E^a_i E^b_j}{\sqrt{|q|}} \; F^k_{ab} + \dots
\end{equation*}
where $N$ is the lapse, $F^k_{ab}$ is the curvature of the connection, $q$ is
the determinant of the three-metric, and the dots indicate the Lorentzian part
of the constraint in which we are not interested at this point (upon symmetry reduction
it is proportional to the first half of the constraint). Most important is the 
presence of the $\frac{E^a_i E^b_j}{\sqrt{|q|}}$ term which leads
to the $\sgn(p) \sqrt{p}$ in the symmetry reduced action, but involves inverse
triad operators in the full theory. Thus as for the inverse volume this
term is represented using a commutator between the volume operator
and holonomies. We will see that for higher $J$ this terms acquires quantum
modifications for small volume analogously to the inverse volume operator.

The resulting operator is given by
\begin{equation}\label{Hop12}
	^{(\half)}\widehat{H}_{GR} = \frac{2 i}{\kappa l_p^2 \gamma^3 \mu_0^3}
	\sum_{ijk} \epsilon^{ijk} \;\; ^{(\half)}\tr \left(
	\hat{h}_i \hat{h}_j \hat{h}_i^{-1} \hat{h}_j^{-1}
	\hat{h}_k \left[ \hat{h}_k^{-1} , \hat{V} \right] \right) \,.
\end{equation}
where we have explicitly indicated that the trace is performed in the fundamental
representation. The curvature term of the classical constraint ($c^2$) has been
regulated with holonomies taken around a closed loop 
(the $\hat{h}_i \hat{h}_j \hat{h}_i^{-1} \hat{h}_j^{-1}$ term).
The $\sgn(p) \sqrt{p}$ term is now represented as 
$\hat{h}_k \left[ \hat{h}_k^{-1} , \hat{V} \right] $ in the constraint operator.
Several key remarks are to be made. First, the holonomy
length parameter $\mu_0$ has been introduced explicitly
with the term in the denominator and  in the holonomy
operators. Second, a factor ordering has been chosen in that the 
$ \sgn(p) \sqrt{p}$ term has been ordered to the right
of the curvature term. This ordering is crucial in the singularity removal
mechanism as we shall see. However, through this choice in
ordering, the constraint is not self-adjoint. Appropriate
self-adjoint constraints  built from
(\ref{Hop12}) have been proposed \cite{Willis:thesis, Bojowald:2004zf}
which have been shown to be non-singular. In this paper
we will consider the non self-adjoint spin $J$ constraint though we note
that self-adjoint constraints can be constructed in an
analogous way which does not effect the results presented here.

Using this formula we can now determine the action of the constraint
operator on the basis states $\bas$. Using the formulas for the holonomies
in the fundamental representation (\ref{hops}) we
get
\begin{equation*}
	^{(\half)}\widehat{H}_{GR} = \frac{48 i}{\kappa l_p^2 \gamma^3
	\mu_0^3} \, \sn^2 \; \cs^2 \; \left[ \sn \,\widehat{V}\,
	\cs- \cs\, \widehat{V} \,\sn \right]
\end{equation*}
whose action on the basis states is
\begin{equation}\label{Hbas}
	^{(\half)}\widehat{H}_{GR}\bas= \frac{3}{2 \kappa l_p^2 \gamma^3
	\mu_0^3} \; \big( V_{\mu+\mu_0} - V_{\mu-\mu_0}\big) \;
	\Big[ \ket{\mu+4\mu_0} -2 \ket{\mu} + \ket{\mu-4\mu_0}
	\Big] \, .
\end{equation}

To understand the singularity removal mechanism we need to consider
physical wave functions which are those annihilated by the constraint
operator. Such states can be expanded using the basis states
as $\ket{\psi} = \sum_{\mu} \, \psi_{\mu} \bas$. Imposing 
the constraint equation leads to a difference equation for the coefficients
$\psi_{\mu}$
\begin{eqnarray}\label{diffeqn12}
	\frac{3}{2 \kappa l_p^2 \gamma^3 \mu_0^3} 
	\Big[ \big( V_{\mu\!+\!5\mu_0} - V_{\mu\!+\!3\mu_0} \big)
	\psi_{\mu+4\mu_0} &-& 
	2  \big( V_{\mu\!+\!\mu_0} - V_{\mu\!-\!\mu_0} \big) \psi_{\mu} 
	\nonumber \\
	&+&  \big( V_{\mu\!-\!3\mu_0} - V_{\mu\!-\!5\mu_0} \big)
	\psi_{\mu-4\mu_0} \Big] = -\widehat{H}_{m} (\mu) \, \psi_{\mu}	
\end{eqnarray}
where we have assumed that the matter constraint $\widehat{H}_{m}$
acts diagonally on the basis states. The key fact about singularity
removal is that the wave function coefficient $\psi_0$ corresponding
to the value of the wave functions at the classical singularity decouples
from the difference equation. This can be seen from the fact that the 
$\big( V_{\mu\!+\!\mu_0} - V_{\mu\!-\!\mu_0} \big)$ factor
that always accompanies $\psi_0$ on the lhs of the difference equation
vanishes. The singularity is then absent provided the matter side of the constraint
annihilates the $\ket{\mu=0}$ state. This has been shown to be
true for various forms of matter including a scalar field.

Let us understand the singularity removal mechanism on a more
general level. The key observation is that the constraint operator {\em
annihilates} the $\ket{\mu=0}$ state since
$\big( V_{\mu\!+\!\mu_0} - V_{\mu\!-\!\mu_0} \big)$  vanishes
identically for $\mu=0$ in equation (\ref{Hbas}). Let us assume in general that the constraint
operator acts by raising and lowering the basis states
\begin{equation*}
	\widehat{H} \,\bas = \sum_{k=-\sigma}^{\sigma}
	\alpha_{\mu}^k \;\ket{\mu+k}
\end{equation*}
where $\sigma$ and $\alpha_{\mu}^k$ are parameters depending on
the details of the operator. Assuming this,
the difference equations for the components $\psi_{\mu}$ will
be
\begin{equation*}
	 \sum_{k=-\sigma}^{\sigma} \alpha_{\mu-k}^k \; \psi_{\mu-k}
	=0
\end{equation*} 
for all values of $\mu$. The statement that the constraint operator
annihilates the $\ket{\mu=0}$ state implies that $\alpha_{0}^k = 0$. In
the difference equation the $\psi_0$ component is always accompanied
with the $\alpha_{0}^k$ term and thus decouples. Had the
$ \sgn(p) \sqrt{p}$ been ordered to the left of the curvature term,
the constraint operator would not annihilate the $\ket{\mu=0}$
state and the singularity would not decouple.

\section{Holonomy Representations}
We would now like to generalize the calculation of the Hamiltonian constraint
operator (\ref{Hop12}) and the inverse volume operator to arbitrary irreducible
representations labeled by spin $J$. That this ambiguity is possible in the full theory was first
elaborated in \cite{Gaul:2000ba}. Of particular interest is what effective semi-classical
equations of motion can be inferred from the constraint operator and
what phenomenological differences can arise from the modifications. After
constructing the arbitrary spin $J$ constraint operator and inverse volume
operator, we will give an explicit construction of the $J=1$
constraint operator and the resulting difference equation will be investigated.
Following that, 
we will propose effective classical equations of motion and bounds
on their validity.

\subsection{Quantum Constraint Operator and Inverse Volume Operator}
Let us propose the generalization of the formula of (\ref{Hop12}) and show that
it is a proper representation of the classical Hamiltonian constraint (\ref{Hclass}).
The formula is
\begin{equation}\label{HJclass}
	^{(J)}H_{GR} = -\frac{3}{\kappa^2 \gamma^3 \mu_o^3 J(J+1)(2J+1)}
	\sum_{ijk}\epsilon^{ijk} \;\; ^{(J)}\tr \Big(
	h_i h_j h_i^{-1} h_j^{-1}
	h_k \left\{ h_k^{-1} , V \right\} \Big) \,.
\end{equation}
We can restore the classical constraint by considering the limit as
$\mu_0$ goes to zero. The holonomies are given by
\begin{equation}\label{happrox}
	h_i = e^{\mu_0 c \tau_i} = 1 + \mu_0 c \,\tau_i +
	\frac{(\mu_0 c)^2}{2} \tau_i^2 + \mathcal{O}(\mu_0^3)
\end{equation}
from which we find
\begin{eqnarray*}
	\sum_{ij} \epsilon^{ijk} h_i h_j h_i^{-1} h_j^{-1}
	&=& 2 \, (\mu_0 c)^2 \tau_k + \mathcal{O}(\mu_0^3) \nonumber \\
	h_k \left\{ h_k^{-1} , V \right\} &=&  -\mu_0 \{ c,V\}\, \tau_k
	+ \mathcal{O}(\mu_0^2) = -\half \kappa \gamma \mu_0
	\;\sgn(p) \sqrt{|p|} \,\tau_k + \mathcal{O}(\mu_0^2) \,.
\end{eqnarray*}
Now using the formula $^{(J)}\tr(\tau_i \tau_j) = -\frac{1}{3} J(J+1)(2J+1) \delta_{ij}$
we get
\begin{equation*}
	\sum_{ijk}\epsilon^{ijk} \; ^{(J)} \tr \left(
	h_i h_j h_i^{-1} h_j^{-1}
	h_k \left\{ h_k^{-1} , V \right\} \right) = \mu_0^3 \gamma \kappa
	J(J+1)(2J+1) \; \sgn(p) \sqrt{|p|}  c^2 + \mathcal{O}(\mu_0^2) 	
\end{equation*}
from which we recover the classical expression given in (\ref{Hclass}).

As in the previous section the classical expression of the Hamiltonian
constraint is promoted to a quantum operator using the holonomy operators
and promoting the Poisson bracket to a commutator. The resulting expression
is given by
\begin{eqnarray}
	^{(J)}\widehat{H}_{GR} &= &
	\frac{3i}{\kappa l_p^2 \gamma^3 \mu_o^3 J(J+1)(2J+1)}
	\sum_{ijk}\!\epsilon^{ijk} \; ^{(J)}\tr \left(
	\hat{h}_i \hat{h}_j \hat{h}_i^{-1} \hat{h}_j^{-1}
	\hat{h}_k \left[ \hat{h}_k^{-1} , \hat{V} \right] \right) \nonumber \\
	&=& \frac{9i}{\kappa l_p^2 \gamma^3 \mu_o^3 J(J+1)(2J+1)}
	\;^{(J)}\tr \Big\{\!\! \left(
	\hat{h}_1 \hat{h}_2 \hat{h}_1^{-1} \hat{h}_2^{-1}  \!-\! 
	\hat{h}_2 \hat{h}_1 \hat{h}_2^{-1} \hat{h}_1^{-1} \right)
	\hat{h}_3 \left[ \hat{h}_3^{-1} , \hat{V} \right]  \!\!\Big\}
\end{eqnarray}
where we have used the fact that the operator is gauge invariant to arrive
at the second line. To calculate the action of the operator we need expressions
for the matrix elements of the holonomies in the spin $J$ representations in
order to perform the trace.
Fortunately such formulas do exist and a particular 
form we will use is
\begin{subequations}\label{holonomies}
\begin{eqnarray}
	(\hat{h}_1)_{mn} &=& T_{mn}\; \sum_{s = |m-n|}^{2J-|m+n|}
	\frac{(-i)^s}{Y_{mn\;s}} \;\;\cs^{2J-s} \;\; \sn^s \label{h1}\\
	(\hat{h}_2)_{mn} &=& T_{mn}\; \sum_{s = |m-n|}^{2J-|m+n|}
	\frac{(-i)^{n-m+s}}{Y_{mn\;s}} \;\; \cs^{2J-s} \;\; \sn^s \label{h2}\\
	(\hat{h}_3)_{mn} &=& e^{im \mu_0 \hat{c}} \; \delta_{mn} \label{h3}
\end{eqnarray}
\end{subequations}
where the matrix elements $m, n \in [-J,J]$ and $T_{mn}$ and
$Y_{mn\;s}$ are constant coefficients given by
\begin{eqnarray}\label{TY}
	T_{mn} &=& \sqrt{(J+m)!\,(J-m)!\, (J+n)!\,(J-n)!} \nonumber \\
	Y_{mn\;s} &=& [J+\half(m+n-s)]!\;  [J-\half(m+n+s)]!\;
	[\half(m-n+s)]!\; [\half(n-m+s)]! 
\end{eqnarray}
The index $s$ in the sum increments by two which implies that if $(m-n)$ is an
even number, the sum over $s$ will comprise even values and vice versa.
We can see that the holonomy operators are more complicated sums and 
products of the basic sine and cosine operators whose action is given
in equations (\ref{hops}). Hence, the holonomies have a well-defined
action on the basis states $\bas$.

With the formula for the holonomies (\ref{holonomies}) we now can determine the
action of the constraint operator. The $\hat{h}_3 \left[ \hat{h}_3^{-1} , \hat{V} \right] $
term is easiest to formulate given the simple expression (\ref{h3}) for
$\hat{h}_3$. We find
\begin{eqnarray*}
	\left( \hat{h}_3 \left[ \hat{h}_3^{-1} , \hat{V} \right]  \right)_{mn}
	&=& \widehat{V} \; \delta_{mn} - e^{im \mu_0 \hat{c}}\,
	\delta_{mo} \,\widehat{V}\, e^{-i o \mu_0 \hat{c}}\, \delta_{on} 
	\nonumber \\
	&=& \left[ \widehat{V} -  e^{im \mu_0 \hat{c}} \,\widehat{V}\,
	e^{-im  \mu_0 \hat{c}} \right] \delta_{mn} \,.
\end{eqnarray*}
This operator acts diagonally on the basis states $\bas$ owing to the action
of $e^{-im  \mu_0 \hat{c}} \bas = \ket{\mu - 2m \mu_0}$. The eigenvalues
are given by
\begin{equation}\label{spop}
	\left( \hat{h}_3 \left[ \hat{h}_3^{-1} , \hat{V} \right]  \right)_{mn}
	\bas = \left[ V_{\mu} -  V_{\mu-2m \mu_0} \right] \delta_{mn} \bas	
\end{equation}
and so the SU(2) matrix elements of this operator are diagonal.

Since $\hat{h}_3 \left[ \hat{h}_3^{-1} , \hat{V} \right] $ has diagonal
matrix elements we need only consider the diagonal
elements $\left(\hat{h}_1 \hat{h}_2 \hat{h}_1^{-1} \hat{h}_2^{-1}  - 
\hat{h}_2 \hat{h}_1 \hat{h}_2^{-1} \hat{h}_1^{-1} \right)_{mm}$
when performing the trace. Using the holonomy formulas (\ref{h1},\ref{h2})
we get
\begin{eqnarray*}
	\left(\hat{h}_1 \hat{h}_2 \hat{h}_1^{-1} \hat{h}_2^{-1}  - 
	\hat{h}_2 \hat{h}_1 \hat{h}_2^{-1} \hat{h}_1^{-1} \right)_{mm}	&=&
	\sum_{n,o,p = -J}^{J}  T_{mn} T_{no} T_{op} T_{pm} 
	\sum_{
		\renewcommand{\arraystretch}{.5}
		\begin{array}{l}
			\scriptstyle{s1 = |m-n| }, \\
			\scriptstyle{s2 = |n-o| }
		\end{array}
	}^{
		\renewcommand{\arraystretch}{.5}
		\begin{array}{l}
			\scriptstyle{2J - |m+n| }, \\
			\scriptstyle{2J- |n+o| }
		\end{array}
	}
	\sum_{
		\renewcommand{\arraystretch}{.5}
		\begin{array}{l}
			\scriptstyle{s1 = |o-p| }, \\
			\scriptstyle{s2 = |p-m| }
		\end{array}
	}^{
		\renewcommand{\arraystretch}{.5}
		\begin{array}{l}
			\scriptstyle{2J - |o+p| }, \\
			\scriptstyle{2J- |p+m| }
		\end{array}
	} \nonumber \\
	\times (-i)^S (-1)^{s3+s4}\!\!\! &&\!\!\! 
	\frac{\big[  (-i)^{o-n+m-p} - (-i)^{n-m+p-o}\big]}
	{Y_{m n \;s1} \,Y_{n o\; s2}\, Y_{o p \;s3}\, Y_{p m\; s4} }
	\; \cs^{8J-S} \; \sn^{S} \nonumber \\
	&=& \sum_{S=0}^{8J} \; Z^S_m \; \cs^{8J-S} \; \sn^{S}
\end{eqnarray*}
where the parameter $S$ is simply $S = s1+s2+s3+s4$ and we have hidden
the complexity of the rhs of the first line in the parameters $Z_m^S$ which
are just constant coefficients. Taken together and performing the trace 
the constraint operator is given by
\begin{equation}\label{Htemp}
	^{(J)}\widehat{H}_{GR} = 
	\frac{9i}{\kappa l_p^2 \gamma^3 \mu_o^3 J(J+1)(2J+1)}	
	\sum_{m=-J}^{J} 
	\sum_{S=0}^{8J} \; Z^S_m \; \cs^{8J-S} \; \sn^{S}
	\left[ \widehat{V} -  e^{im \mu_0 \hat{c}} \widehat{V}
	e^{-im  \mu_0 \hat{c}} \right] 
\end{equation}

We can simplify this formula greatly by looking in detail
at the coefficients $Z_m^S$ many of which vanish. Let us
again display the formula
\begin{equation}\label{Z}
	Z_m^S = 	\sum_{n,o,p = -J}^{J}  T_{mn} T_{no} T_{op} T_{pm}
	\sum_{
		\renewcommand{\arraystretch}{.5}
		\begin{array}{l}
			\scriptstyle{s1 = |m-n| }, \\
			\scriptstyle{s2 = |n-o| }
		\end{array}
	}^{
		\renewcommand{\arraystretch}{.5}
		\begin{array}{l}
			\scriptstyle{2J - |m+n| }, \\
			\scriptstyle{2J- |n+o| }
		\end{array}
	}
	\sum_{
		\renewcommand{\arraystretch}{.5}
		\begin{array}{l}
			\scriptstyle{s1 = |o-p| }, \\
			\scriptstyle{s2 = |p-m| }
		\end{array}
	}^{
		\renewcommand{\arraystretch}{.5}
		\begin{array}{l}
			\scriptstyle{2J - |o+p| }, \\
			\scriptstyle{2J- |p+m| }
		\end{array}
	}  (-i)^S (-1)^{s3+s4}
	\frac{\big[  (-i)^{o-n+m-p} - (-i)^{n-m+p-o}\big]}
	{Y_{m n \; s1}\, Y_{n o \; s2}\, Y_{o p \;s3}\, Y_{p m s4} }
\end{equation}
with the restriction that $s1+s2+s3+s4 = S$. To simplify this we will use two
important facts. First, note that when $(o-n)+(m-p)$ in the sum is an even number,
the numerator vanishes. Second, as we have stated since the s parameters increment
by two, their evenness is determined by the parameters $m,n,o,p$, thus for instance
if $m-n$ is even then $s1$ is even.

We now show that the coefficients satisfy the following:
\begin{subequations}
\begin{eqnarray}
	Z_m^S &=& 0 \;\;\;\; \text{for $S$ odd}  \label{Zodd}\\
	Z_m^0 &=& Z_m^{8J} = 0 \label{Z0}\\
	Z_m^S &=&  - Z_{-m}^S \label{Zflip}\,.
\end{eqnarray}
\end{subequations}
To show (\ref{Zodd}) we note that if $S \equiv s1+s2+s3+s4$ is odd
then this implies that $(m-n)+(n-o)+(o-p)+(p-m)$ is also odd. However,
this quantity is equal to zero and hence even. Thus $S$ must be even and
$Z_m^S=0$ for $S$ odd. (\ref{Z0}) can be shown by calculating $Z_m^0$
explicitly. If $S=0$ then this implies that all $s_i$  are zero. This
only occurs in the sum over $n,o,p$ when $n=o=p=m$. Hence, 
$o-n+m-p=0$ which is even and thus the numerator vanishes. A
similar calculation holds for $Z_m^{8J}$. To show (\ref{Zflip})
one can make the substitution in (\ref{Z}) $m,n,o,p \rightarrow -m, -n,
-o, -p$. Using the fact that $T_{mn} = T_{-m -n}$ and
$Y_{m n\; s} = Y_{-m -n \;s}$ (which follow from the definitions (\ref{TY}))
it is trivial to show that $Z_m^S =  - Z_{-m}^S$.

Using these properties we simplify the constraint operator
to 
\begin{equation}\label{HJop}
	^{(J)}\widehat{H}_{GR} = 
	\frac{-9i}{\kappa l_p^2 \gamma^3 \mu_0^3 J(J+1)(2J+1)}	
	\sum_{m=-J}^{J}
	\sum_{S'=1}^{4J-1} \; Z^{2S'}_m \; \cs^{8J-2S'} \; \sn^{2S'} \;
 	  e^{im \mu_0 \hat{c}} \,  \widehat{V} \,
	e^{-im  \mu_0 \hat{c}} 	
\end{equation}
where the term in (\ref{Htemp}) involving only $\widehat{V}$ vanishes
due to (\ref{Zflip}). The action of this operator on the basis states
$\bas$ can be understood as follows. We showed that the
term $e^{im \mu_0 \hat{c}} \widehat{V} e^{-im  \mu_0 \hat{c}}$
acts diagonally on the basis states and in the next section we will show
that it approximates the $ \sgn(p) \sqrt{p}$ term in the classical constraint. 
The $ \cs$
and $\sn$ terms each raise and lower the basis states
by discrete steps up to $\pm 8J$. Since the sum is over
even powers of $ \cs$ and $\sn$, $\widehat{H}$ will
raise and lower the basis states by even values of $\mu_0$.
Thus in general we can say the action on the basis states
is
\begin{equation*}
	^{(J)}\widehat{H}_{GR} \; \bas = 
	\sum_{k=-4J}^{4J} \alpha^k_{\mu} \;
	\ket{\mu+2 k \mu_0} 	
\end{equation*}
where $\alpha^k_{\mu}$ are coefficients determined by equation
(\ref{HJop}). The resulting difference equation will be of order
$8J$. Furthermore, the fundamental step size of the difference
equation is $\delta_{\mu} = 2 \mu_0$. Note that for the
$J=1/2$ operator some further cancellation occurs and
the difference equation given in equation (\ref{diffeqn12})
has step size equal to $4 \mu_0$. While for arbitrary $J$ similar
cancellations may occur, we will show that for $J=1$ the
step size is indeed $2 \mu_0$ and the difference equation is of
order $8J = 8$. We can state for certain that the smallest possible
step size is $2 \mu_0$.

With regards to the singularity behavior we have stated that singularity
removal occurs if the constraint operator annihilates the $\ket{\mu=0}$
state. To show that this occurs for arbitrary $J$ we will use
the fact that $Z_m^S =  - Z_{-m}^S$. The constraint
will thus contain the following term
\begin{equation*}
	\sum_{m=1}^J Z^{2S'}_m \left(e^{im \mu_0 \hat{c}} \widehat{V}
	e^{-im  \mu_0 \hat{c}} - e^{-im \mu_0 \hat{c}} \widehat{V}
	e^{im  \mu_0 \hat{c}} \right)	
\end{equation*}
whose action on the basis states contributes a
$\sum_{m=1}^{2S'} \left( V_{\mu-2m \mu_0} - V_{\mu+2m \mu_0}
\right)$
term. It is easy to see that this vanishes for $\mu=0$ and hence
the singularity decouples from the difference equation.

Turning now to the inverse volume operator, the formula
for arbitrary $J$ is \cite{Bojowald:2002ny}
\begin{eqnarray*}
	\widehat{V^{-1}} &=& 
	\Big[ - \frac{4i}{ l_p^2 \gamma \mu_0 J(J+1)(2J+1)}
	\sum_i \tr \left( \tau_i \hat{h}_i [\hat{h}_i^{-1} , \hat{V}^{1/2}]
	\right) \Big]^6 \nonumber \\
	&=&\Big[ - \frac{12i}{ l_p^2 \gamma \mu_0 J(J+1)(2J+1)}
	\sum_i \tr \left( \tau_3 \hat{h}_3 [\hat{h}_3^{-1} , \hat{V}^{1/2}]
	\right) \Big]^6 
\end{eqnarray*}
where again we have used gauge invariance to arrive at the
second line. Using the formula for the holonomy (\ref{h3}) we find
through similar calculations to those of the
 $\hat{h}_3 \left[ \hat{h}_3^{-1} , \hat{V} \right] $ term of the
constraint operator, the inverse volume operator acts diagonally
on the basis states with eigenvalues denoted by $d_J$
\begin{equation}\label{dJ}
	d_J(\mu) = \Big[ \frac{12}{ l_p^2 \gamma \mu_0 J(J+1)(2J+1)}
	\sum_{m=-J}^J \;m\; V^{1/2}_{\mu+2m \mu_0} \Big]^6
\end{equation}
and can be approximated for large $J$ by the function
\begin{eqnarray}
	d_J(a) &\approx& \frac{1}{a^3} \; D^6(a^2/a_*^2) \nonumber \\
	D(q) &=& 2 q^{1/4} \Big[ \frac{4}{7}
	\left[ (q+1)^{7/4} + \sgn(q-1) |q-1|^{7/4} \right]
	-\frac{16}{77} \left[ (q+1)^{11/4} +  |q-1|^{11/4} \right]
\end{eqnarray}
where $a_* = \sqrt{\frac{\gamma J \mu_0}{3}}l_p$ is the characteristic
scale factor below which the quantum corrections
are large. The eigenvalues are bounded with the maximum value
occurring near $a_*$. For $a \gg a_*$ the function $d_J(a)$ approximates
well the classical expression $a^{-3}$. Below $a_*$,  $d_J(a)$
behaves polynomially and can be approximated by treating the sum
as an integral to get
\begin{equation}\label{djsmall}
	d_J(a) \approx \left( \frac{12}{7} \right)^6 \frac{a^{12}}{a_*^{15}} \,.
\end{equation}
Our goal was to note explicitly the role played by the regularization length $\mu_0$.
We note that the effect is simply to shift $a_*$ to larger or smaller volumes.

We have thus succeeded in formulating the constraint operator (\ref{HJop}) and 
inverse volume operator (\ref{dJ}) for arbitrary $J$. The key results
for the Hamiltonian constraint are that the difference equation is higher order
for larger $J$, yet the step size remains on the same order. The singularity removal
mechanism remains valid for arbitrary $J$. For the inverse volume operator, the
eigenvalues $d_J$ are bounded near a critical scale factor $a_*$ whose value
scales both with $\mu_0$ and $J$. We do not
explore any further the general properties of the quantum constraint for arbitrary
$J$ as we do not have a simple formula for the coefficients $Z^S_m$ and
the resulting difference equation would be quite complicated
and none the enlightening.

\subsection{Quantum Theory for J = 1}
We now wish to calculate an explicit example of the quantum
operator given in equation $(\ref{HJop})$. For simplicity we
choose the $J=1$ representation. We will derive the 
difference equation and show that under appropriate conditions
the solutions approximately satisfy the Wheeler-DeWitt differential
equation. We will also examine the local stability of the
difference equation to determine the behavior of spurious
solutions.

After a little labor the $J=1$ operator is given by
\begin{equation*}
	^{(1)}\widehat{H}_{GR} = \frac{12}{\kappa l_p^2 \gamma^3 \mu_0^3}
	\left[ \,\cs^6 \sn^2 - 2  \cs^4 \sn^4
	-  \cs^2 \sn^6 \,  \right]
	\left[ e^{i\mu_0 \hat{c}} \hat{V}  e^{-i\mu_0 \hat{c}} -
	e^{-i\mu_0 \hat{c}} \hat{V}  e^{i\mu_0 \hat{c}} \right]
\end{equation*}
whence the action on the basis states is
\begin{eqnarray*}
	^{(1)}\widehat{H}_{GR} \bas = - 	
	\frac{3}{32 \kappa  \gamma^2 \mu_0^2}\;
	s_1(\mu) &\Bigg\{& \ket{\mu+8 \mu_0}
	-4  \ket{\mu+6\mu_0} -4 \ket{\mu + 4 \mu_0} \nonumber \\
	&+&4\ket{\mu+2\mu_0} +6 \ket{\mu} +4 \ket{\mu-2 \mu_0} \\
	&-&4 \ket{\mu-4 \mu_0} -4 \ket{\mu-6 \mu_0}+ \ket{\mu-8 \mu_0}
	\Bigg\} \nonumber
\end{eqnarray*}
where we have defined
\begin{equation*}
	s_1(\mu) \equiv \frac{V_{\mu+2\mu_0} - V_{\mu-2\mu_0}}
	{\gamma l_p^2 \mu_0}
\end{equation*}
and it can easily be shown that for large volume
$s_1(p) \approx \sqrt{p}$.
The resulting difference equation is
\begin{eqnarray}\label{diffeqn1}
	\frac{-3}{32 \kappa  \gamma^2 \mu_0^2} &\Bigg\{& \!\!\!
	\cof{+8\mu_0} - 4 \cof{+ 6 \mu_0} -4 \cof{+ 4 \mu_0} \nonumber \\
	\!\!&+&\!\! 4 \cof{+ 2\mu_0} + 6\cof{} +4 \cof{-2\mu_0} \\
	\!\!&-&\!\! 4 \cof{-4\mu_0} -4 \cof{-6 \mu_0} + \cof{-8\mu_0} \Bigg\}
	= -\widehat{H}_{m} (\mu) \,\psi_{\mu} \,. \nonumber 
\end{eqnarray}
As predicted in the general case, the difference equation is of order $8J=8$
with step size $\delta \mu = 2 \mu_0$.

To make contact with the Wheeler-DeWitt equation we will
make the following assumptions. First is that we are in the large volume
limit where $\mu \gg \mu_0$. Second is that
the wave function coefficients $\psi_{\mu}$ do not vary sufficiently fast on
the order of $\delta\mu = 2\mu_0$ such that we can approximate
the discrete coefficients with a continuous function $\psi(\mu)$. Whether
or not this assumption holds true for large volume is model dependent
and an example where this continuum approximation fails is the isotropic
flat model with a positive cosmological constant. The consequences for that
model are discussed in depth in \cite{LQC_Hphys}. Yet, let us
assume for the moment that we can make these assumptions, hence
we can Taylor expand the function $\psi(\mu+\delta) \approx \psi(\mu)
+\frac{d \psi}{d \mu} \delta + \half \frac{d^2 \psi}{d \mu^2}\,\delta^2$.
Under these assumptions $\psi(\mu)$ will approximately satisfy a differential
equation.

The Wheeler-DeWitt equation is derived from the classical action
(\ref{Hclass}) be performing the usual Schrodinger quantization
where the operator $\hat{c}$ is quantized to $-\frac{i}{3} \hbar \kappa
\gamma \frac{\partial}{\partial p}$. The resulting
differential equation is
\begin{equation}\label{WDeqn}
	\frac{\kappa \hbar^2}{3} \frac{\partial^2}{\partial p^2}
	\Big[ \sqrt{p} \,\psi(p) \Big] + \hat{H}_{m}
	\psi(p) = 0 \,.
\end{equation}
Turning to the difference equation let us define $t_{\mu} \equiv s_1(\mu) \psi_{\mu}$
and the difference equation simplifies to 
\begin{eqnarray*}
	- \frac{3}{32 \kappa  \gamma^2 \mu_0^2} &\Bigg\{&
	t_{\mu+8\mu_0}-4  t_{\mu+6\mu_0} -4 t_{\mu+4\mu_0}
	+4  t_{\mu+2 \mu_0} +6 t_{\mu} +4t_{\mu-2\mu_0} \\
	&-&4 t_{\mu-4\mu_0}-4 t_{\mu-6\mu_0} +t_{\mu-8\mu_0}
	\Bigg\} = -\widehat{H}_{m} (\mu)\, \psi_{\mu} \,.
\end{eqnarray*}
Using $p = \frac{\mu \gamma l_p^2}{6}$ and Taylor expanding
$t(p)$ we find
\begin{eqnarray*}
  && t_{\mu+8\mu_0}-4  t_{\mu+6\mu_0} -4 t_{\mu+4\mu_0} +4  t_{\mu+2 \mu_0} 
  \nonumber \\
  &+&6 t_{\mu} +4t_{\mu-2\mu_0} -4 t_{\mu-4\mu_0}-4 t_{\mu-6\mu_0} +t_{\mu-8\mu_0}
  \approx - \frac{32}{9} \mu_0^2 \gamma^2 l_p^4 \; \frac{\partial^2\, t(p)}{\partial\, p^2}
  \,.
\end{eqnarray*}
For large volume $t(p) \equiv s_1(p)\, \psi(p) \approx \sqrt{p} \,\psi(p)$ 
and plugging this
in we find that the difference equation is approximated
by the Wheeler-DeWitt equation.

Next we would like to examine the local stability of the difference equation.
Since the difference equation is of higher order than the second order
Wheeler-DeWitt equation, we must determine the behavior of the spurious
solutions. As determined in \cite{Bojowald:2003dn}, if the higher order
difference equation admits solutions with amplitudes that grow
(locally), then the difference equation is not locally stable and
this might call into the question the validity of the quantization
as any semi-classical solutions would quickly become dominated
by the expanding spurious solutions.

Let us be more precise. We will consider the large volume limit
in a regime where the matter contribution is small. We assume that the volume
is large enough such that the variations of $V_{\mu}$ and
$H_{m}(\mu)$ are small. We can thus approximate the difference equation
by one with constant coefficients
\begin{eqnarray}\label{diffconst}
  \psi_{\mu+8\mu_0}&-&4  \psi_{\mu+6\mu_0} -4 \psi_{\mu+4\mu_0} +4  \psi_{\mu+2 \mu_0}
  + \psi_{\mu} +4 \psi_{\mu-2\mu_0}
  \nonumber \\
  &-&4 \psi_{\mu-4\mu_0}-4 \psi_{\mu-6\mu_0} +\psi_{\mu-8\mu_0}
  = \frac{32 \kappa \gamma^2 \mu_0^2 H_{m}(\mu)}{3 s_1(\mu)} \psi_{\mu}
  =P \, \psi_{\mu}
\end{eqnarray}
where we assume that on the order of $\delta \mu = 8 \mu_0$ that
$P \equiv \frac{32 \kappa \gamma^2 \mu_0^2 H_{m}(\mu)}{3 s_1(\mu)}$ is
constant. The difference equation with constant coefficients can be solved exactly
by assuming solutions of the form $\psi_{\mu} = A \,z^{\mu/\mu_0}$ where
$z \in \C$. The condition of local stability states that we
should consider solutions of the homogeneous equation
(with $P=0$) and that the difference equation is
locally stable if all solutions have norm equal
to one, that is $|z|^2=1$ \cite{Bojowald:2003dn}. This will guarantee
that spurious solutions which don't approximate a semi-classical
Wheeler-DeWitt solution will not come to dominate.

\begin{figure}[ht]
\begin{center}
  \includegraphics[width=11cm, keepaspectratio]
        {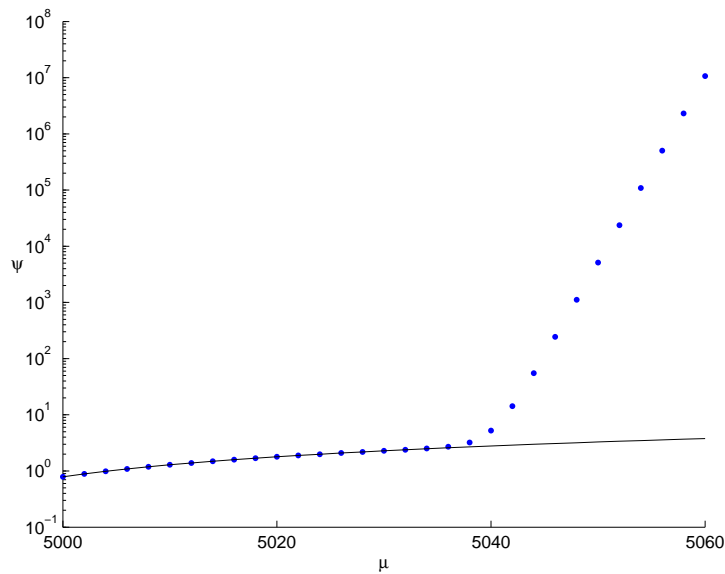}
\end{center}
\caption{Log plot of the difference equation solution (dots)
compared with Wheeler-DeWitt solution (solid line) for
a massless scalar field at large volume. The presence of spurious
solutions to the higher order difference equation eventually dominate
to solution.
}
\label{j1instab}
\end{figure}

Plugging in our ansatz $\psi_{\mu} = A \,z^{\mu/\mu_0}$ to the homogeneous
difference equation (\ref{diffconst}) we find the following condition on
$z$
\begin{equation*}
  z^8 - 4z^6 - 4 z^4 + 4 z^2+6 +4 z^{-2} -4 z^{-4} -4 z^{-6}
  +z^{-8} = 0 \,.
\end{equation*}
A numerical calculation of the roots shows that there are solutions with
norms greater than one, thus the $J=1$ difference equation is {\em not}
locally stable. The result of this behavior is plotted
in figure (\ref{j1instab}) for a massless scalar field with
constant momentum where $H_m(\mu) = \half d_J(\mu) P_{\phi}^2$. The
figure shows a solution to the difference equation (\ref{diffeqn1}) and
a solution to the Wheeler-DeWitt equation (\ref{WDeqn}) for comparison.
Initial conditions are specified on $\psi_{\mu}$ to match the Wheeler-DeWitt
solution and the solution is evolved forward using the difference
equation. It is clear that the solution of the difference equation
follows the Wheeler-DeWitt solution briefly, after which the
spurious non semi-classical solutions quickly dominate the wave function.

The presence of the ill behaved spurious solutions does not
by itself represent a problem with the quantization. To truly determine
the physical consequence requires a detailed understanding of the physical
inner product. With a physical inner product, one could identify
which solutions are physical based on the notion of probability, that
is unphysical solutions would have either vanishing or infinite
physical norm and would be modded out of the physical Hilbert
space. An exampled where this happens is in the symmetry reduced
Plebanski model of \cite{LQC_Hphys}. There, extra solutions
have zero physical norm and the resulting physical Hilbert space
is one dimensional for matter in the form of a cosmological
constant. If the spurious solutions here do indeed have
zero physical norm then the quantization presented would
have the correct semi-classical limit. If not this would
indicate the need for different quantizations or even the possibility
that the higher spin quantizations do not represent consistent quantizations
of isotropic LQC. This might indicate a preference of the quantum
theory toward the use of the fundamental representation for
defining the Hamiltonian constraint. There are possible indications
that this behavior also occurs in the full theory where a large
set of spurious solutions exist for the higher $J$ gravitational
constraint operator \cite{Alex_private}.  

\subsection{Effective Classical Theory}
We now turn to the question of what sort of semi-classical
equations of motion can be inferred from the constraint operator given
in equation (\ref{HJop}). While in general the effective equations of motion
should be derived from the wave function solutions themselves (for instance
by finding semi-classical states), here we will take the simplest route. 
Our motivation comes from the path integral quantization of LQC
where a discretized path integral involves integration
over a classical action
(for details on the path integral derivation see \cite{LQC_Hphys}).
A detailed derivation of the path integral shows that the 
effective classical Hamiltonian constraint is given
by the p-q symbol $H_{eff} = \frac{ \braketfull{c}{\hat{H}}{p} }{ \braket{c}{p}}$.
Since the Hamiltonian constraint of LQC consists of a self-adjoint part
(the $h_i h_j h_i^{-1} h_j^{-1}$ term) of which $\ket{c}$ is an eigenstate,
and a part ($h_k \left\{ h_k^{-1} , V \right\}$) of which $\ket{p}$ is an eigenstate
we need only consider the eigenvalues of the two operators to
get an effective classical constraint written in terms of a classical
$c$ and $p$.  We
note that a more detailed consideration involving semi-classical
states could lead to additional corrections. Equations of
motion can then be derived from the effective Hamiltonian
through the Hamiltonian equations
$\dot{p} = \{p,H_{eff}\}$ and $\dot{c} = \{c,H_{eff}\}$. We are interested
in determining what modifications arise for instance to the Friedmann
equations from the quantum effects.

We now show that two kinds of corrections arise in the effective constraint. The first
occurs for small volumes and arises from the  $h_k \left\{ h_k^{-1} , V \right\}$
term which we will show behaves as a modification to the
classical $\sgn(p)\sqrt{|p|}$ term in the constraint. The second
modification is due to the curvature term $h_i h_j h_i^{-1} h_j^{-1}$ and
amounts to a modification of the classical $c^2$ term. This modification
is evident when the connection $c$ is large or equivalently (for flat models)
when the extrinsic
curvature is large. We will show that for large volume and
small extrinsic curvature, the correct classical equations are recovered.

\begin{figure}[ht]
\begin{center}
  \includegraphics[width=11cm, keepaspectratio]
        {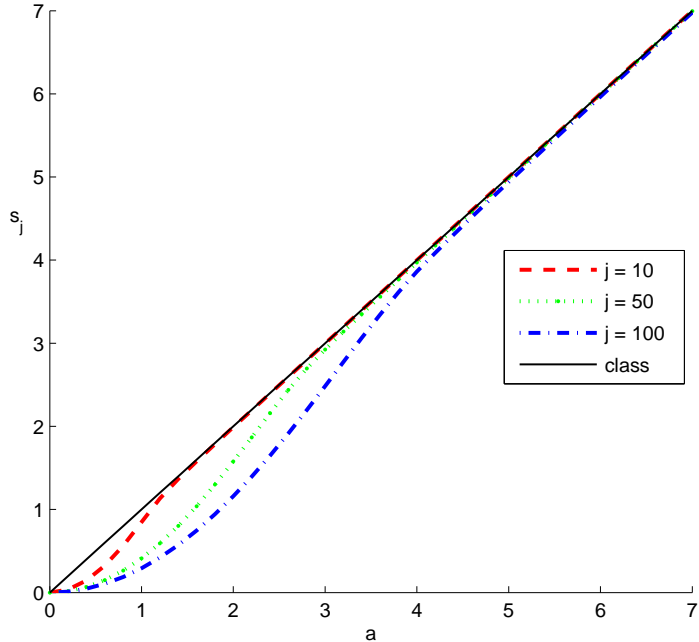}
\end{center}
\caption{LQC modified function $s_J(a)$ (\ref{sjeigs}) compared with
the classical expression for the function $\sgn(p) \sqrt{|p|} = a$. The
modified $s_J(a)$ function behaves quadratically for
$a < a_*  = \sqrt{\frac{\gamma J \mu_0}{3}} l_p$. The values
of $a_*$ in the graph are $1.290, 2.89, 4.08$
for $J= 10, 50, 100$ respectively.
}
\label{sjfig}
\end{figure}

Let us now calculate the corrections explicitly. From equation (\ref{spop})
we get the formula for the $h_3 \left\{ h_3^{-1} , V \right\}$ term
as
\begin{equation*}
	\left( \hat{h}_3 \left[ \hat{h}_3^{-1} , \hat{V} \right]  \right)_{mn}
	\bas = \left[ V_{\mu} -  V_{\mu-2m \mu_0} \right] \delta_{mn}	\,.
\end{equation*}
Next the curvature term $h_i h_j h_i^{-1} h_j^{-1}$ can be calculated
by expanding the holonomies $h_i = e^{\mu_0 c \tau_i}$ as done
for equation (\ref{happrox}). If taken to order $(\mu_0 c)^4$ we find that
\begin{equation*}
	h_1 h_2 h_1^{-1} h_2^{-1} -h_2 h_1 h_2^{-1} h_1^{-1}	
	= 2 (\mu_0 c)^2 \tau_3 + (\mu_0 c)^3 (\tau_1-\tau_2) -
	\frac{2}{3} (\mu_0 c)^4 \tau_3 + \mathcal{O}\big((\mu_0 c)^5 \big) \,.
\end{equation*}
Now using $(\tau_3)_{mn} = im \delta_{mn}$ and the fact that $(\tau_1)_{mn}$
and $(\tau_2)_{mn}$ are off-diagonal matrices we find after performing the
trace
\begin{eqnarray}\label{Heff}
	H_{eff} &=& \frac{-9i}{\kappa \gamma^3 l_p^2 \mu_0^3 J(J+1)(2J+1)}
	\left( 2 \mu_0^2 c^2 - \frac{2}{3}\mu_0^4 c^4 + \mathcal{O}(\mu_0^5 c^5)
	\right) \sum_{m=-J}^J im [V_{\mu}-V_{\mu-2 m \mu_0}]
	\nonumber \\
	&=& -\frac{3}{\kappa \gamma^2}\; s_J \; \left( c^2 - \frac{1}{3}\mu_0^2 c^4
	\right) + \mathcal{O}\Big(\frac{\mu_0^3 c^5}{\kappa \gamma^2}\Big)
\end{eqnarray}
and this is the formula we will use as our effective constraint. Notice the
similarity to the classical constraint in (\ref{Hclass}) except for two modifications.
The first is the presence of the function $s_J$ which we have defined to be
\begin{equation}\label{sjeigs}
	s_J \equiv - \frac{6}{\gamma l_p^2 \mu_0 J(J+1)(2J+1)}
	\sum_{m=-J}^J m V_{\mu-2 m \mu_0}
\end{equation}
which we now show approximates  $\sgn(p) \sqrt{|p|}$. This can
be seen by noting that for large volume $V_{\mu-2 m \mu_0} \approx
V_{\mu} - m \gamma \mu_0 l_p^2 / 2 \;\sgn(p) \sqrt{|p|} +
\mathcal{O}(\frac{\mu_0^2 \gamma^2 l_p^4}{\sqrt{p}})$. Thus
we get
\begin{eqnarray*}
	s_J &\approx& - \frac{6}{\gamma l_p^2 \mu_0 J(J+1)(2J+1)}
	\sum_{m=-J}^J m \left( V_{\mu} - \frac{m \gamma \mu_0 l_p^2}{2}
	\sgn(p) \sqrt{|p|} + \mathcal{O}(\frac{\mu_0^2 \gamma^2 l_p^4}{\sqrt{p}})
	\right) \nonumber \\
	&\approx& \sgn(p) \sqrt{|p|} + \mathcal{O}
	\Big(\frac{\mu_0 \gamma l_p^2}{\sqrt{p}} \Big)
\end{eqnarray*}
where we have used $\sum_{m=-J}^{J} m^2 = 1/3\, J(J+1)(2J+1)$ to arrive
at the second line. As for the formula for $d_J(\mu)$ we can approximate
$s_J(\mu)$ for large values of $J$ by treating the sum as an integral. The 
resulting formula is
\begin{eqnarray*}
	s_J(\mu) \approx \sqrt{\frac{\gamma l_p}{6}}
	\frac{1}{2 \mu_0^3 J^3} \Big\{ \frac{J \mu_0}{5}
	\left[ (\mu+2J\mu_0)^{5/2} + \sgn(\mu-2J\mu_0)
	|\mu-2J\mu_0|^{5/2} \right] \nonumber \\
	-\frac{1}{35} \left[ (\mu+2J\mu_0)^{7/2}
	- |\mu-2J\mu_0|^{7/2} \right] \Big\}
\end{eqnarray*}
which implies that written in terms of the scale factor $a$
\begin{eqnarray}
	s_J(a) &=& a \,S(a^2/a_*^2) \nonumber \\
	S(q) &=& \frac{4}{\sqrt{q}} \left\{
	\frac{1}{10} \left[ (q+1)^{5/2} + \sgn(q-1)|q-1|^{5/2}
	\right] - \frac{1}{35} \left[ (q+1)^{7/2}
	- |q-1|^{7/2} \right] \right\}
\end{eqnarray}
where as for $d_J(a)$ we have  the {\em same} critical scale
factor $a_*=\sqrt{\frac{\gamma J \mu_0}{3}} l_p$
below which quantum corrections occur. The function
$S(q)$ for $q > 1$ is approximately equal to one which
implies that for $a > a_*$, $s_J(a) \approx a = \sqrt{|p|}$.
For small volume $S(q)$ behaves as $\frac{6}{5} \sqrt{q}$ hence
$s_J(a)$ behaves quadratically with a
\begin{equation}\label{sjsmall}
	s_J(a) \approx \frac{6}{5} \frac{a^2}{a_*} \;\;\;\;\; a \ll a_* \,.
\end{equation}
A plot of the function $s_J(a)$ compared with $\sqrt{p} = a$ is
shown in figure (\ref{sjfig}). It is clear that $s_J$ 
behaves quadratically for small $a$ and changes behavior near
$a_*$ after which it approximates the classical expression $\sqrt{p}$.

Thus we have determined the effective Hamiltonian constraint 
given in equation (\ref{Heff}) which is valid in the regime
where the connection component $c$ is small or more precisely
when $\mu_0 c \ll 1$. The function $s_J(\mu)$ in turn approximates
the classical $\sgn(p) \sqrt{|p|}$ term for large volume when
$\mu \gg 2 J \mu_0$ or equivalently when $a \gg a_*$.

Let us now make contact with the standard formulation
of isotropic cosmology written in terms of the scale factor $a$.
We assume that the orientation of the triad is positive ($p, \mu > 0$)
so we can drop the absolute value and sign terms in the formulas.
According to equations (\ref{p}, \ref{pmu}) we have the relation
between $\mu$ and the scale factor as $a^2 = \frac{\gamma}{6} \mu l_p^2$.
From the Hamiltonian equations and the vanishing of the constraint
we can derive the Friedmann equation. We first have
\begin{eqnarray*}
	\dot{p} &=& \{ p, H_{eff} \} = -\frac{1}{3} \kappa \gamma 
	\frac{\partial H_{eff}}{\partial c} \nonumber \\
	&=& \frac{1}{\gamma} \big( 2 c - \frac{4}{3} \mu_0^2 c^3 \big)
	s_J \,.
\end{eqnarray*}
Using $p = a^2$ we can use this equation to get the Friedmann
equation in terms of the connection as
\begin{equation*}
	\left( \frac{\dot{a}}{a} \right)^2 = 
	\frac{s_J^2 c^2}{\gamma^2 a^4} \left(1-\frac{4}{3} \mu_0^2 c^2 
	+\frac{4}{9} \mu_0^4 c^4 \right) \,.
\end{equation*}
From this it is clear that the criteria $\mu_0 c \ll 1$ corresponds
to a criteria on the scale factor velocity being small $\mu_0 \gamma \frac{a}{s_J} \dot{a} \ll 1$.
To get this equation in standard form we need to solve for the
connection in terms of the matter Hamiltonian. We can get this
from the vanishing of the full Hamiltonian constraint $H = H_{eff}+H_{m}=0$.
This gives
\begin{equation*}
	c^2 = \frac{\kappa \gamma^2}{3 s_J} H_{m} +
		\frac{\kappa^2 \gamma^4 \mu_0^2}{27 s_J^2} H_{m}^2
\end{equation*}	
where we have only kept the terms to second order in $H_{m}$. Putting these
two together we get the modified Friedmann equation
\begin{eqnarray}\label{Fried}
	\left( \frac{\dot{a}}{a} \right)^2 &=&
        \frac{\kappa}{3} \; \frac{s_J}{a^4} \;  H_{m}
        - \frac{\kappa^2 \gamma^2 \mu_0^2 }{9 a^4} \;H_{m}^2
        \nonumber \\
        &=& \frac{\kappa}{3}\; S(a)\; \rho_{m}
        - \frac{1}{9} \kappa^2 \gamma^2 \mu_0^2 a^2 \;
        \rho_{m}^2	
\end{eqnarray}
where the matter density is simply $\rho_{m} \equiv a^{-3} H_{m}$
and as before $s_J(a)=  a\, S(a)$. The modified Friedmann equation
remains valid provided
$\mu_0 \gamma \frac{a}{s_J} \dot{a} \ll 1$. This in turn
places a bound on the matter density $\rho_{m} \ll
\frac{3 S(a)}{\kappa \mu_0^2 \gamma^2 a^2}$ below which
the effective equations remain valid. From the modified 
Friedmann equation it is clear that in the large
volume regime (where $S(a) \approx 1$) and small curvature
(where we neglect the $\rho_{m}^2$ term) we recover the standard
Friedmann equation $\left( \frac{\dot{a}}{a} \right)^2 =  \frac{\kappa}{3} \rho_{m}$.

Let us reiterate the main results of the section. The modified
effective Hamiltonian constraint is given in equation (\ref{Heff})
and we rewrite it here
\begin{equation}
	H_{eff} = -\frac{3}{\kappa \gamma^2} s_J \left( c^2 - \frac{1}{3}\mu_0^2 c^4
        \right)
\end{equation}
where the function $s_J$ is given in equation (\ref{sjeigs}) and approximates
$\sgn(p) \sqrt{|p|}$ for large volume. The modified Friedmann dynamics
given in equation (\ref{Fried}) was calculated through the Hamiltonian
equations derived from the effective constraint. The effective 
constraint is valid when $\mu_0 c \ll 1$ which in turn
places a bound of validity determined by both the extrinsic curvature and on the matter
density.

\section{Discussion}
In this paper we have sought to clear up certain issues in LQC, namely the complete
consequences of quantizing the Hamiltonian constraint using higher spin representations for
the holonomies. In addition we have fixed the constants necessary to make contact on a consistent basis with the
standard action of GR. We have found that the use of higher spin holonomies to regulate the
gravitational part of the constraint operator leads to modifications of the $\frac{E^a_i E^b_j}{\det(|q|)}$
term. The modifications are qualitatively similar to those of the inverse scale factor which
lead to the $d_J (a)$ function. This is not unexpected since we are quantizing the constraint
operator to take into account factors involving the inverse triad (the $1/det(|q|)$ term). This
modification, leading to the $s_J (a)$ function, is another important consequence of a key mantra
of LQC: to remain as close to the full theory of loop quantum gravity as possible. We could
have exploited the symmetry of the isotropic model and quantized the $\sgn(p) \sqrt{|p|}$
term avoiding any inverse triad operators. Because this is not possible in the full theory it is not
done in LQC and the non-perturbative corrections are imported from the full theory.

We have been guided by certain simplicity considerations. Previous studies have used
the $J = 1/2$ gravitational part of the constraint (and the resulting difference equation) while
freely choosing an arbitrary representation to define the inverse volume operator in the
matter part of the constraint. While there is certain freedom to specify the representations
differently and quantize the two parts in a different manner, 
for instance to define a $J_G$ for the gravitational part and a $J_{m}$ 
for the matter part, the simplest choice is to use the same
representation. This ambiguity also exists in the parameter
$\mu_0$ appearing in the matter and gravitational part.   

Thus it is important to reexamine the phenomenological consequences in light of the
modifications to the Friedmann equation (\ref{Fried}) that arise for arbitrary values of $J$ . 
Let us first concentrate on the modifications arising from $s_J (a)$. 
In the instance of a massless scalar field
 we have for the Friedmann equation
\begin{equation*}
	\left( \frac{\dot{a}}{a} \right)^2 = \frac{\kappa}{3} \frac{s_J(a)}{a^4}
	\half d_J(a) P_{\phi}^2 
\end{equation*}
where we have used $H_{m} = \half d_J(a) P_{\phi}^2$ with $P_{\phi}$ being
constant. To understand the behavior of the scalar field let us
de-parameterize the equations of motion to remove the notion
of time. This is in accord with notions of quantum gravity where
coordinate time does not have any physical meaning. Thus we
are interested in how the scalar field evolves with the scale
factor as opposed to coordinate time. 
From the Hamiltonian equations we find that
$\dot{\phi} = d_J(a) P_{\phi}$ and we de-parameterize
as $d \phi / da = \dot{\phi}/ \dot{a}$ to get
\begin{equation*}
	\frac{d\phi}{da} = \sqrt{\frac{6 a \, d_J(a)}{\kappa \,S(a)}}
\end{equation*} 
which implies that the field is pushed to {\em higher} values
in the region below $a_*$ since the $S(a)$ appearing in the
denominator is suppressed in that region. The addition of the $s_J$ corrections
can therefore improve the viability of LQC to push the inflaton up its potential
to set the initial conditions for slow-roll inflation.

The consequences of the modifications to the Friedmann equation quadratic in matter have
not been explored. Important though is that we have given a precise bound
$\mu_0 \gamma \frac{a}{s_J} \dot{a} \ll 1$
on the validity of the first order effective constraint. This is in contrast with the criteria for
validity cited in other phenomenological investigations $a/\dot{a} \gg \sqrt{\gamma} l_p$
\cite{Bojowald:2004xq} based on the notion that the Hubble length should not be smaller than the fundamental length of discreteness $\sqrt{\gamma} l_p$. The criteria given here can be understood heuristically as follows. The discreteness
of the difference equation plays a role precisely where the Wheeler-DeWitt approximation
(\ref{WDeqn}) breaks down, that is when the wave function changes significantly on the order of the
fundamental step size $\delta \mu = 2 \mu_0$. 
If we consider a Wheeler-DeWitt solution that is locally
oscillatory  $\psi(\mu) = exp(i 2 \pi \mu / \lambda)$ where $\lambda$ is the wavelength, then the discreteness becomes
important when $\lambda \approx 4 \mu_0$. Given that in the Wheeler-DeWitt equation
$\hat{c} = -\frac{i \gamma l_p^2}{3} \partial/ \partial p$ we
find that the connection $c$ is related to $\lambda$ as $c = 4 \pi / \lambda$.
 Using this we recover the criteria
that the higher order corrections occur when $\mu_0 c \approx 1$. We can therefore
understand that these corrections are a direct result of discreteness effects of the
difference equation.

The term quadratic in the matter density is negative definite in the Friedmann equation
and hence can act as an effective compactness even in the non-compact model considered in this paper. Whether or not this would indeed imply a re-collapse depends on the
inclusion of the full corrections to arbitrary order of $\mu_0 c$. Furthermore, classical descriptions
in this regime are not expected to be valid since, as stated in the previous paragraph, this
corresponds to the regime where the discreteness is dominant. A model where this occurs is
deSitter space which is discussed in detail in \cite{LQC_Hphys}. In that model classically the extrinsic curvature grows with time and the wave functions become non semi-classical around a critical volume given by $V_c = \left(\frac{6}{\Lambda \gamma^2 \mu_0^2} \right)^{3/2}$
which is precisely when $\mu_0 c = \pi$. Above the critical volume, the wave function decays rapidly indicating a classically forbidden region and thus an
effective compactness. Thus this correction can have drastic consequences for the evolution
of the universe. For ordinary matter such as dust or radiation, the matter density drops off
sufficiently fast for these corrections not to be relevant at large volumes. Yet the fact that
observations suggest that the universe might be in an asymptotically deSitter space call into
question the validity of LQC for large volumes and a more detailed
derivation of LQC from the full theory would likely solve this problem.

We are in the process of examining numerically the full phenomenological consequences of the modified
Friedmann equation (\ref{Fried}) for more complicated forms of matter. 
The fact that the $s_J$ corrections help push the inflaton up its
potential is a benefit to this scenario and would help enlarge
the parameter space that leads to successful inflation.

The key parameter involved in the quantum corrections is the regulating length $\mu_0$ . A
deep understand of the origin of this parameter is essential before exact phenomenological
predictions can be made. In the full theory one takes the limit as $\mu_0$ goes to zero and
a well-defined operator remains which due to diffeomorphism invariance is independent of
the regulator \cite{Thiemann:1996aw}. In LQC the corrections for $s_J$, $d_J$ as well as for large extrinsic curvature
depend explicitly on the value of $\mu_0$ . If we were to take the limit of small $\mu_0$, the phenomenological consequences of the quantum corrections would be pushed into the deep Planckian
regime. This might wash out any possible observational signals from LQC. Yet, a smaller
value of $\mu_0$ would also push to larger volumes the validity of LQC for the deSitter example
discussed above. In \cite{Ashtekar:2003hd} it is argued heuristically that $\mu_0$ is fixed to a value of
$\sqrt{3}/4$ in
the Hamiltonian constraint based on the smallest allowed area of the full theory. A more
thorough understand of the parameter $\mu_0$ and its derivation from the full theory is required
to settle this issue.

A key issue with which we have not dealt is the need to derive the effective equations
of motion from solutions of the difference equation itself. Several proposals have been put forth 
to date on this matter. In one \cite{Willis:thesis}, kinematical 
semi-classical states are constructed and the effective
constraint is calculated from the expectation value of the constraint operator.
The validity of this is not entirely clear since the semi-classical states are not physical
states in that they are not annihilated by the constraint operator. The corrections
to the constraint do include a $\mu_0^2 c^4$ term as we showed in this paper (in that paper the
corrections are calculated for large volume for the $J = 1/2$ constraint so would not include
$s_J$ corrections). The exact factors for these corrections do not agree however. In addition
corrections arise in that model due to the spread of the wave function, something of which we
have not considered here in deriving the effective equations of motion. 

In another technique
\cite{Date:2004zd}, WKB type solutions of the difference equation are calculated and 
from that an effective constraint is extracted. In this paper the modified constraint includes a 
$-\frac{3}{2 \gamma^2 \kappa} \left[ s_{1/2}(\mu+4\mu_0) +s_{1/2}(\mu-4\mu_0)
\right] c^2$ term which to leading order is equivalent to the effective constraint
of equation (\ref{Heff}) for $J=1/2$. More detailed calculations in
the WKB context \cite{Banerjee:2005ga} show the higher order
corrections in $\mu_0 c$ appearing in our effective constraint
(\ref{Heff}). In that paper the effective constraint for large volume is given
by $H_{eff} = - \frac{3}{\kappa \gamma^2 \mu_0^2} \sqrt{p} \sin^2(\mu_0 c)$
(where we have adjusted the multiplicative factors to agree with ours)
which is precisely the effective constraint presented in this paper
had we calculated all the higher order corrections for the $J=1/2$
constraint (the $c^2 - \frac{1}{3}\mu_0^2 c^4$ term can be seen as the
small $\mu_0 c$ expansion of $\sin^2(\mu_0 c)/\mu_0^2$).
In addition the WKB effective constraint contains a potential term which
contributes for small volume. 
Again the interpretation of dynamics in the WKB setting is not entirely
clear especially  in the setting of the discrete difference equation of LQC.
In addition the approximation used in \cite{Date:2004zd} relies on correlating wave function 
solutions of the difference
equation on scales smaller than the fundamental step size $\delta \mu =  4 \mu_0$. 
It can be shown in LQC
that quantum interference only occurs for quantum states defined at values of $\mu$ differing
by an integer times the fundamental step size \cite{LQC_Hphys}. 
Thus, quantum mechanically LQC does
not correlate the quantum wave function at a given volume $\mu$ and a neighboring value 
$\mu+ \epsilon$ where $\epsilon < \delta \mu$.

A further method to determine semi-classical dynamics involves
introducing a coordinate time parameter with which to
consider evolution of Gaussian wave packets \cite{Bojowald:2004zf}. In
that paper the wave packets are shown to follow the classical trajectory
even to small volumes. The relevance of the solutions to
physical solutions (annihilated by the constraint operator and
independent of any time parameterization) has yet to be determined.

Each of the techniques to derive effective equations of motion are performed with models
with zero degrees of freedom, and thus none consider physical semi-classical states. In
quantum gravity where coordinate time is not a physical degree of freedom, one needs to
include higher degrees of freedom and choose one of them to play the role of a clock. The
model of a massless scalar field provides a testing ground for these ideas. In this model,
the volume is a monotonically increasing function (classically) and can play the role of a
clock. A semi-classical state peaked around some value of the scalar field at a given volume
can be evolved forward in "time" with the difference equation and the resulting trajectory
compared with the effective equations of motion. This would have the benefit that the wave
functions considered would be physical states annihilated by the constraint operator. Early
results for the spin $1/2$ difference equation indicate that the first order corrections for the
effective equations match well with the results from the difference equations to volumes near
the Planck scale. While this issue would be trickier to study for the higher order spin $J$
difference equation, the agreement for the spin $1/2$ constraint is indicative of the validity of
the effective equations.

\begin{acknowledgments}
We would like to thank Abhay Ashtekar, Martin Bojowald, and
Alejandro Perez for important comments. We thank Parampreet Singh
and Ghanashyam Date for stimulating conversations pertaining to
this work. This work has been supported by NSF grants
PHY-0354932 and PHY-0090091 and the Eberly Research
Funds of Penn State University.
\end{acknowledgments}


\end{document}